\documentclass[11pt]{article}
\usepackage{moriond,epsfig}

\def\Journal#1#2#3#4{{#1} {\bf #2}, #3 (#4)}

\def\NIM{\em Nucl. Instrum. Methods}

\def\PLB{{\em Phys. Lett.}  B}
\def\PRL{\em Phys. Rev. Lett.}
\def\PRD{{\em Phys. Rev.} D}

\def\be{\begin{equation}}
\def\ee{\end{equation}}
\def\bea{\begin{eqnarray}}
\def\eea{\end{eqnarray}}

\newcommand{\ifm}[1]{\relax\ifmmode #1\else $#1$\hskip 0.15cm\fi}

\newcommand{\tbar}{\ifm{\overline{t}}}
\newcommand{\pbar}{\ifm{\overline{p}}}
\newcommand{\ppb}{\ifm{p\pbar}}
\newcommand{\ttb}{\ifm{t\tbar}}

\begin{document}
\vspace*{4cm}
\title{Top quark and $W/Z$ results from the Tevatron}

\author{Dhiman Chakraborty \\ (representing the D\O\ and CDF Collaborations)}

\address{Department of Physics, Northern Illinois University,\\
DeKalb, IL 60115, USA}

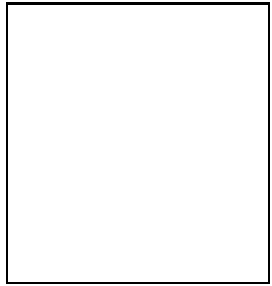
\maketitle\abstracts{
We summarize results of some crucial measurements of the top quark and $W/Z$ 
boson properties carried out by the D\O\ and the CDF collaborations at the 
Tevatron collider at Fermilab based on data collected during Run 1 (1992-96). 
Among the interesting properties measured are the pair-production cross 
section and the mass of the top quark, and the mass and the width of the $W$ 
boson. Searches for singly produced top quarks and for certain non-standard 
production and decays of the top quark, as well as studies of angular 
correlations in the production and decay of the top quarks are also presented.
Expectations from the ongoing Run 2 of the Tevatron, presently in its 
second year, are discussed.
}

\section{Introduction}\label{sec:introduction}
Studies of the top quark and the $W$ and $Z$ bosons provide testing grounds 
for many important properties and calibrations of the Standard Model (SM)
at large mass scales through their production rates, kinematic distributions, 
and decay characteristics.
In the SM, $m_t$ and $m_W$ constrain $m_H$.
Hence, precision measurements of these help guide the search for 
the SM Higgs boson.
Measurement of the $W$ width provides a stringent test of the SM and
helps to constrain certain scenarios beyond the SM.

The large mass of the top quark sets it apart from all other fermions. 
With a lifetime of \mbox{$\sim$ $10^{-24}$ s}, the top quark is expected to 
decay before hadronization.
This gives us an opportunity to study a bare quark, free from the long-range 
effects of the strong interaction, such as color confinement.
Studies of the top quark could also shed light on the mechanism of mass
generation and mass-dependent couplings.
Significant deviations from SM predictions in mass, width, decay 
characteristics, and kinematic distributions could lead to new physics.
Furthermore, top quark decay is an excellent place to look for on-shell 
production of certain particles beyond the SM (e.g.,  $\tilde t$, $H^\pm$, 
\ldots) that are believed to be heavier than other SM particles.

\section{The collider and the detectors}\label{sec:col_dets}
Run 1 of the Tevatron (1992-96) delivered $125\pm 6$  pb$^{-1}$ of data
consisting of over $5\times 10^{12}$ $p\bar p$ collisions at a center-of-mass
energy of 1.8 TeV.
With $\sigma(p\bar p\rightarrow t\bar tX)\approx 6$ pb for 
$m_t\approx 175$ GeV, and $\sigma(p\bar p\rightarrow WX)\approx 24$ nb,
$\sim$750 $t\bar t$ events and $\sim$$3\times 10^6$ $W$ events are expected to 
have been produced. 
However, these events must be detected and filtered out from a much larger
number of other processes with similar signatures.
The two detectors, D\O, and CDF, are both modern 
multipurpose detectors designed to do just that.~\cite{tevdets}
They consist of vertex detectors, precision tracking chambers, finely 
segmented hermetic calorimeters, muon momentum spectrometers, and fast
data acquisition systems with several levels of online triggers and filters.
\section{Studies of the top quark}\label{sec:top}
\subsection{Pair production cross section and mass measurements}
At the Tevatron \ppb collider, most top quarks are pair-produced via the strong
interaction through an intermediate gluon.
This is the production channel that led to the joint discovery of the top 
quark by D\O\ and CDF in 1995.~\cite{tt_obs}
In the SM, the top quark should decay almost exclusively to a $W$ boson and a 
$b$ quark. Therefore, the final state of a \ttb decay is characterized by the
decay modes of the two $W$ bosons: dilepton (BR$=\frac{1}{9}$), lepton+jets
(BR$=\frac{4}{9}$), and all-jets (BR$=\frac{4}{9}$). 
Tagging of the $b$ quark jets using semileptonic decay modes or by isolating 
secondary decay vertices is a powerful means of suppressing background.
All of these decay channels have been used in various studies of the properties
of the top quark, including the \ttb production cross section of 
$5.7 \pm 1.6$ pb by the D\O\ collaboration~\cite{cstt_d0} and 
$6.5^{+1.7}_{-1.4}$ pb by the CDF collaboration,~\cite{cstt_cdf} and of the 
top quark mass of $174.3 \pm 5.1$ GeV jointly by the two 
collaborations.~\cite{mt}
\subsection{Search for single top production}
A second production mode is predicted to exist, where top quarks are produced
singly through an electroweak $Wtb$ vertex.
Measurement of the electroweak production of single top quarks could provide
the magnitude of the CKM matrix element $V_{tb}$,
since the cross section is proportional to $|V_{tb}|^2$.
The SM predictions for single top production cross section are 0.73 pb in the
$s$ channel, and 1.73 pb in the $t$ channel. 
In addition to having a smaller cross section than that for pair production, 
single top production suffers from larger background contamination.
Neither experiment saw conclusive evidence of signal, but were able to place
upper limits on the production cross section. At 95\% CL, D\O\ puts an upper
limit of 17 pb on the $s$ channel and 22 pb on the $t$ channel cross section 
while the corresponding CDF limits are 13 pb and 18 pb 
respectively.~\cite{single_top} 
CDF also puts a limit of 14 pb on the combined cross section.
\subsection{$W$ helicity in top decays}
At the leading order, the fraction of longitudinally polarized $W$ bosons
in top decays is given by
\be
{\cal F}_0 \equiv \frac{\Gamma(h_W=0)}{\Gamma(h_W=0)+\Gamma(h_W=-1)}
=  \frac{m_t^2/(2m_W^2)}{1+m_t^2/(2m_W^2)} \approx 0.70.
\ee
The large mass of the top quark exposes the longitudinal mode of the $W$ boson,
possibly offering a window to electroweak symmetry breaking.
The charged lepton from a $W_-$ ($W_0$) tends to move opposite (perpendicular)
to the $W$ direction of motion.
Consequently, the polarization of the $W$ is reflected in the $p_T$
distribution of the $e$ or $\mu$ in its leptonic decays.
Fits of \mbox{$h_W=0$, $h_W=-1$, and $h_W=+1$} Monte Carlo to CDF dilepton
and lepton+jets data yield
${\cal F}_0(W) = 0.91\pm0.37\pm0.13$, ${\cal F}_+(W) = 0.11\pm0.15\pm0.06$,
consistent with SM predictions of $\sim$0.70 and 0 
respectively.~\cite{w_helicity}

\subsection{Top-antitop spin correlations}
The top quark lifetime is much smaller than the timescale for
hadronization which subsequently would lead to spin decorrelation.
The decay products of top quarks produced in a definite spin state should
therefore display angular correlations that characterize the production 
process.
In the optimal basis, if the angle between a negatively (positively) charged 
lepton or $d$ ($\bar d$) -type quark and the spin quantization axis is denoted
by $\theta_-$ ($\theta_+$), then the differential decay rate of the top quark
can be parametrized as~\cite{ttspin_theory}
\be
\frac{1}{\sigma} \frac{d^2\sigma}{d(\cos \theta_+) d(\cos \theta_-)}
= \frac{1 + \kappa \cos \theta_+ \cos \theta_-}{4}; \qquad -1<\kappa<1.
\ee
All information on spin correlation is contained in $\kappa$.

For \ppb collisions at $\sqrt{s}=1.8$ TeV, $q\bar{q}$ annihilation
through the $s$-channel via a spin-1 gluon is expected to account for 
$\sim$90\% of the \ttb production cross section. 
This leads to an expectation of $\kappa \approx 0.9$.
An analysis of the D\O\ dilepton data consisting of 6 candidate events favors
a positive value for $\kappa$ 
($\kappa>-0.25$ at 68\% CL).~\cite{ttspin_result}
\subsection{\ttb invariant mass and kinematics}
Both CDF and D\O\ have searched for non-SM top quark condensates in the \ttb
invariant mass distributions. 
Comparing a theoretical predictions for 
$\sigma(q\bar q\rightarrow Z^\prime)B(Z^\prime\rightarrow \ttb)$ as functions 
of $m_{Z^\prime}$ with upper limits on the former based on data, CDF puts
an lower limits on the mass of a leptophobic $Z^\prime$ boson in a model of
topcolor-assisted technicolor: 
$m_{Z^\prime}>480$ GeV at 95\% CL for $\Gamma_{Z^\prime}=0.012m_{Z^\prime}$ and
$m_{Z^\prime}>780$ GeV at 95\% CL for $\Gamma_{Z^\prime}=0.04m_{Z^\prime}$.~\cite{m_zprime}
Preliminary results from D\O\ are consistent with this finding.

Studies of kinematics of \ttb production and decays are important tests of
QCD and electroweak theories.
Both experiments have also examined distributions of various kinematic 
quantities for the candidate events.~\cite{tt_kinematics}
These include transverse momenta and rapidity of individual top quarks and of
the \ttb system, opening angles etc. 
All distributions were found to be in excellent agreement with SM predictions.
\subsection{Search for $t\rightarrow H^+b$}
Charged Higgs arise in the simplest extension of the SM Higgs sector to a
two-Higgs doublet model.
If $m_{H^+}<m_t-m_b$, then $t\rightarrow H^+b$ could compete with 
$t\rightarrow W^+b$ depending on $m_{H^+}$ and $\tan{\beta}$, the ratio of
vacuum expectation values of the two scalar doublets.
Direct searches for charged Higgs pair production at LEP have resulted in 
a lower limit of $m_{H^+}>69$ GeV at 95\% CL.~\cite{ch_lep}
The decay $t\rightarrow H^+b$ dominates only if $\tan{\beta}$ is either very 
large ($>40$) or very small ($<0.9$) and loses ground as $m_{H^+}$ increases.
Both D\O\ and CDF have searched for the subsequent decay
$H^+\rightarrow \tau\nu$ which is the favored mode if $\tan{\beta}>1$.
In addition, D\O\ has searched the region of small $\tan{\beta}$, where the
favored decay mode would be $H^+\rightarrow c\bar s$, by examining the 
consistency of their lepton+jets data with SM predictions. 
The searches, based on leading order calculations, result in the exclusion of 
most of the $[m_{H^+},\tan{\beta}]$ parameter space where 
$B(t\rightarrow H^+b)>0.5$.

\section{Studies of the $W$ boson}\label{sec:W}
\subsection{$W$ mass measurements}\label{sec:m_W}
CDF has used both $W\rightarrow e\nu$ and $W\rightarrow \mu\nu$ decays to 
obtain $m_W=80.433\pm0.079$ GeV while D\O\ measures $m_W=80.483\pm0.084$ GeV 
using the $W\rightarrow e\nu$ channel only.~\cite{mw_tev}
The two experiments have combined their results and that from the UA2 
experiment to get a hadron collider average of $m_W=80.456\pm0.059$ GeV
which is comparable in precision and in good agreement with the LEP result
of $m_W=80.450\pm0.039$ GeV.~\cite{mw_lep}
The combined world average now stands at $m_W=80.451\pm0.032$ GeV.~\cite{w_tev_comb}
\subsection{Width of the $W$ boson}\label{sec:gamma_W}
Both D\O\ and CDF have measured $\Gamma_W$ by studying the $W$ transverse
mass distribution (the direct method). 
The results are $\Gamma_W=2.04\pm 0.11$ (stat) $\pm 0.09$ (sys) 
GeV (CDF) and $\Gamma_W=2.23^{+0.18}_{-0.17}$ GeV (D\O, preliminary).~\cite{gamma_w}
These results are combined with the width extracted from the ratio of 
$W$ and $Z$ leptonic partial cross sections.
The combined result from the Tevatron is $\Gamma_W=2.160\pm 0.047$ GeV.~\cite{w_tev_comb}
This is the most precise measurement of the $W$ width yet, in good agreement
with the LEP direct measurement of \mbox{$\Gamma_W=2.150\pm0.091$ GeV}.
The preliminary world average is $\Gamma_W=2.158\pm0.042$ GeV.~\cite{w_tev_comb}
\section{Future outlook}\label{sec:future}
Run 2 of the Tevatron, presently in an early stage, holds much potential. 
The accelerator upgrade is designed to increase the center-of-mass energy by
10\% (which translates to a $\sim$30\% increase in 
$\sigma(\ppb\rightarrow\ttb)$), and a much higher integrated luminosity:
2 fb$^{-1}$ per experiment in Run 2a, 15 fb$^{-1}$ in Run 2b.
The detector upgrades will significantly improve signal efficiencies and
background rejection.
The number of events used for extracting the $W$ mass and width measurements by
D\O\ in Run 2 is expected to be $\sim 430$ times that in Run 1.  
A combined enhancement of a factor of $\sim$300 for \ttb events, and even 
greater for single top events are expected by the end of run 2.
These should enable us to reduce the uncertainties on the top and $W$ mass
measurements by a factor of 2 to 3.
All other studies, including those on which first results have been obtained,
are dominated by statistical uncertainties. 
These will benefit greatly from Run 2.
Table \ref{tab:summary} summarizes the current results and expectations in the
near future for some of the studies of top quark physics.
\begin{table}[h]
\caption{Summary of current results and future expectations on some physics
topics related to the top quark.
\label{tab:summary}}
\vspace{0.4cm}
\begin{center}
\begin{tabular}{|l|c|c c c c|} \hline
{ Top quark} & & 
\multicolumn{4}{|c|}{ Precision}\\ \cline{3-6}
 property &   Run 1 measurement & Run 1 
&  Run 2a &  Run2b &  LHC\\ \hline\hline
 Mass  &  $174.3\pm 3.3\pm 3.9$ GeV & 2.9\% 
&  1.2\% &  1.0\% & \\ \hline
 $\sigma(\ttb)$  &  $6.5^{+1.7}_{-1.4}$ pb { (CDF)} & 25\% 
&  10\% &  5\% &  5\%\\
 $\sigma(\ttb)$ &  $5.9 \pm 1.7$ pb { (D\O)}& & & & \\ \hline
 $F_0(W)$ &  $0.91\pm0.37 \pm 0.13$ & 0.4 &  0.09 &  0.04 
&  0.01\\
 $F_+(W)$ &  $0.11\pm0.15 \pm 0.06$ & 0.15 &  0.03 
&  0.01 &  0.003\\ \hline
 $ R \equiv  $
&  $0.94^{+0.31}_{-0.24}$ (3-gen.) & 30\% & 4.5\% &  0.8 \% 
&  0.2\%\\
 $\frac{B(t\rightarrow W^+b)}{B(t \rightarrow Wq)}$&  $>0.61$ at 90\% C.L. 
& & & & \\
\hline
 $ |V_{tb}|$ from \ttb &  $0.96^{+0.16}_{-0.12}$ (3-gen.) & & & &\\
& $>0.051$ at 90\% C.L. & $>0.05$ &  $>0.25$ &  $>0.50$  
&  $>0.90$\\ \hline
 $\sigma$(single top)
&  $<13.5$ pb &  $-$ & 20\% &  8 \% &  5\%\\
 $\Gamma(tWb)$ & $-$ &  $-$ & 25\% &  10 \% &  10 \%\\
 $|V_{tb}|$ from single top& $-$ &  $-$ & 12\% &  5 \% &  5 \%\\ \hline
 $B(t\rightarrow \gamma q)$ &  $<0.03$ (95\% C.L.) &  $<0.03$
&  ? &  ? &  ? \\
 $B(t\rightarrow Z q)$ &  $<0.32$ (95\% C.L.) &  $<0.3$
&  $<0.02$ &  ? &  ? \\
\hline
\end{tabular}
\end{center}
\end{table}

\end{document}
